# Anisotropic Generalization of the ΛCDM Universe Model with Application to the Hubble Tension

Øyvind G. Grøn

Faculty of Technology, Art and Design, Oslo and Akershus University College of Applied Sciences, St. Olavs Plass, P.O. Box 4, NO-0130 Oslo, Norway; oyvind.gron.no@gmail.com

**Abstract:** I deduce an exact and analytic Bianchi type I solution of Einstein's field equations, which generalizes the isotropic ΛCDM universe model to a corresponding model with anisotropic expansion. The main point of the article is to present the anisotropic generalization of the ΛCDM universe model in a way suitable for investigating how anisotropic expansion modifies observable properties of the ΛCDM universe model. Although such generalizations of the isotropic ΛCDM universe model have been considered earlier, they have never been presented in this form before. Several physical properties of the model are pointed out and compared with properties of special cases, such as the isotropic ΛCDM universe model. The solution is then used to investigate the Hubble tension. It has recently been suggested that the cosmic large-scale anisotropy may solve the Hubble tension. I consider those earlier suggestions and find that the formulae of these papers lead to the result that the anisotropy of the cosmic expansion is too small to solve the Hubble tension. Then, I investigate the problem in a new way, using the exact solution of the field equations. This gives the result that the cosmic expansion anisotropy is still too small to solve the Hubble tension in the general Bianchi type I universe with dust and LIVE (Lorentz Invariant Vacuum Energy with a constant energy density, which is represented by the cosmological constant) and anisotropic expansion in all three directions—even if one neglects the constraints coming from the requirement that the anisotropy should be sufficiently small so that it does not have any significant effect upon the results coming from the calculations of the comic nucleosynthesis during the first ten minutes of the universe. If this constraint is taken into account, the cosmic expansion anisotropy is much too small to solve the Hubble tension.

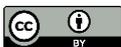

## 1. Introduction

In the present article, I solve Einstein's field equations and calculate exact analytic expressions for the co-moving volume, the Hubble parameter, and the shear scalar as functions of time for the general Bianchi type I universe model with anisotropy in all three directions. The solution is presented in a form suitable for investigating how anisotropic expansion modifies observable properties of corresponding isotropic universe models.

I subsequently apply these solutions to the question of whether the upper limit of the expansion anisotropy of the universe, as restricted by different types of observations, is large enough to permit a solution of the Hubble tension.

The Hubble tension is the fact that early universe measurements and calculations to determine the value of Hubble's constant and late time measurements have given results with a difference that is larger than the uncertainties of the determined values. A recent review of *late* time measurements has been given by Adam Riess and co-workers [1] with the result that the Hubble constant is $H_0$ = 73.04 ± 1.04 km s$^{-1}$ Mpc$^{-1}$. The most recent result using supernovae and quasars was announced on 30 August 2023 by T. Liu et al. [2]. They found $H_0$ = 73.51 ± 0.67 km s$^{-1}$ Mpc$^{-1}$. A review of the *early* universe measurements has been given by the Planck team [3] with the result $H_0$ = (67.4 ± 0.5) km s$^{-1}$ Mpc$^{-1}$. Further references are found in these articles.

A large number of articles have been published with proposed solutions to the Hubble tension, but so far, no solution has been generally accepted. A review [4] with 709 references has recently been published by Maria Dainotti and co-workers.

I here discuss the proposal by V. Yadav [5] that cosmic anisotropy can solve the problem. Ö. Akarsu et al. [6,7] have given constraints on the anisotropy of a Bianchi type I spacetime extension of the standard ΛCDM universe model. They found that the present value of the anisotropy parameter is restricted to $\Omega_{\sigma 0} < 10^{-18}$ (95% C.L.) from CMB+Lense data and that the introduction of spatial curvature or anisotropic expansion, or both, on a generalized ΛCDM universe model, does not offer a possible relaxation to the $H_0$ tension. The results of the present analysis, based upon an exact solution of Einstein's field equations, confirm this result.

In Figure 1 and Table 3 of ref. [5] it is indicated that the Hubble tension is solved in the context of an anisotropic Bianchi type I universe model. However it is not explained where this solution comes from. It is tempting to think that it is the anisotropy of the expansion which solves the Hubble tension. In the present paper, I investigate whether this is the case.

The method which is used here to determine the effect of the anisotropy of the expansion upon the value of the Hubble's constant, i.e., the present value of the Hubble parameter, is the following. First a formula giving the evolution of the Hubble parameter for the considered anisotropic universe model is deduced. Then, the anisotropy is assumed to vanish, and the corresponding formula with vanishing anisotropy is written down. These formulae connect the Hubble parameter at an arbitrary point of time with its value at the present time, i.e., the Hubble constant, for an anisotropic and the corresponding isotropic universe model. The initial value is chosen as the value determined from the Planck observations of the temperature fluctuations of the cosmic microwave background radiation. These measurements determined the value of the Hubble parameter at the point of time 380,000 years after the Big Bang. This value is an observed quantity and is independent of whether the cosmic expansion is assumed to be anisotropic or isotropic. Hence, this is the initial value in both the evolution equation of the Hubble parameter in the anisotropic and the isotropic case. Then, these equations are used to calculate the present values of the Hubble parameter, i.e., the values of the Hubble constant, in an anisotropic- and an isotropic universe. Finally, the difference of the values of the Hubble constant with anisotropic and isotropic expansion is calculated and compared with the difference of the values from early and late time measurements, which makes up the Hubble tension.

## 2. Anisotropic Generalization of the ΛCDM Universe Model

The Bianchi universes are universe models which are homogeneous but not necessarily isotropic. There exist nine Bianchi-type universe models.

The Bianchi type I universe models are the only universe models of the Bianchi type that reduce to flat, isotropic FLRW universe models in the case of vanishing anisotropy. Hence they are candidates for generalizing the ΛCDM universe models to universe models permitting anisotropic expansion.

Concerning the possibility that cosmic expansion anisotropy should be able to solve the Hubble tension, the investigations in the present paper show that the anisotropy of our universe, as described in terms of the most general Bianchitype I universe model, is much too small to be able to solve the Hubble tension. Hence describing the expansion anisotropy in terms of any other type of the Bianchi universe models is not expected to lead to a solution of the Hubble tension.

In particular, the Bianchi type V universe models do not contain the ΛCDM universe models because they are spatially curved. They are not expected to be more realistic universe models than those of the Bianchi type I and will, therefore, not be considered in the present work.

*2.1. Field Equations for the Bianchi Type I Universe Models*

A 55-year-old article [8] by P. T. Saunders deserves to be mentioned as a pioneering work. He considered a general Bianchi type I universe model, allowing for different expansion factors in three orthogonal directions with scale factors $a_1(t)$, $a_2(t)$, $a_3(t)$. Using units with the velocity of light $c = 1$ the line-element has the form

$$ds^2 = -dt^2 + a_1^2(t)dx^2 + a_2^2(t)dy^2 + a_3^2(t)dz^2. \tag{1}$$

The average scale factor, $a(t)$, and co-moving volume, $V(t)$, are

$$a(t) = (a_1 a_2 a_3)^{1/3}, \quad V = a^3 = a_1 a_2 a_3. \tag{2}$$

The scale factors are normalized so that the co-moving volume at the present time is $V(t_0) = 1$. The redshift of observed radiation from a source emitting the radiation at a point of time $t_e$ is

$$z = \frac{1}{V^{1/3}(t_e)} - 1. \tag{3}$$

The directional Hubble parameters and the average Hubble parameter are

$$H_i = \frac{\dot{a}_i}{a_i}, \quad H = \frac{1}{3}(H_1 + H_2 + H_3) = \frac{1}{3}\frac{\dot{V}}{V}. \tag{4}$$

The average deceleration parameter is

$$q = -1 - \frac{\dot{H}}{H^2} = 2 - 3\frac{V\ddot{V}}{\dot{V}^2}. \tag{5}$$

As usual $q > 0$ means deceleration of the cosmic expansion and $q < 0$ means acceleration.

It has also been usual to introduce the shear scalar,

$$\sigma^2 = (1/6)\left[(H_1 - H_2)^2 + (H_2 - H_3)^2 + (H_3 - H_1)^2\right], \tag{6}$$

which is a kinematic quantity representing the anisotropy of the cosmic expansion. Yadav [5] has shown that

$$\sigma = \sigma_0 / V \tag{7}$$

where $\sigma_0$ is the present value of the shear scalar. This relationship is valid independently of the energy and matter contents of the universe. It is a kinematical relationship in the Bianchi type I universe models.

Saunders first considered a Bianchi type I universe filled with cold matter in the form of dust with density $\rho_M$ and vanishing pressure, radiation with density $\rho_R$, and dark energy in the form of LIVE having a constant density, $\rho_\Lambda$, which can be represented by the cosmological constant $\Lambda$.

I will here deduce a relationship first shown by Saunders, following [9], but giving more detailed explanations. With the line element (1) Einstein's field equations

$$R_{\mu\nu} - (1/2)g_{\mu\nu}R = \kappa T_{\mu\nu} \tag{8}$$

can be written in the form [10]

$$(\ln V)^{\cdot\cdot} + H_1^2 + H_2^2 + H_3^2 = \Lambda \tag{9}$$

$$(VH_1)^{\cdot} = (VH_2)^{\cdot} = (VH_3)^{\cdot} = \Lambda V \tag{10}$$

and furthermore, as shown by P. Sarmah and U. D. Goswami [11],

$$\sum_{i=1}^{3} \frac{\ddot{a}_i}{a_i} = -\frac{\kappa}{2}(\rho + 3p) \tag{11}$$

where $\rho$ and $p$ are the sum of the mass-energy densities and pressure, respectively, of the matter, radiation, and dark energy contents of the universe.

Due to Equation (11), it is natural in the Bianchi type I universe models to define an effective gravitational mass density, $\rho_G$, by

$$\sum_{i=1}^{3} \frac{\ddot{a}_i}{a_i} = -\frac{\kappa}{2}\rho_G . \tag{12}$$

Giving

$$\rho_G = \rho + 3p . \tag{13}$$

There is attractive gravity for $\rho_G > 0$ and repulsive gravity for $\rho_G < 0$. Cold matter has vanishing pressure, and radiation has a positive pressure $p_R = (1/3)\rho_R$, so both cause attractive gravity. On the other hand, LIVE has negative pressure $p_\Lambda = -\rho_\Lambda$ giving $\rho_{G\Lambda} = -2\rho_\Lambda < 0$, causing strong repulsive gravity corresponding to twice the gravity of the mass density.

Adding Equation (10) gives

$$V(\dot{H}_1 + \dot{H}_2 + \dot{H}_3) + \dot{V}(H_1 + H_2 + H_3) = 3\Lambda V \tag{14}$$

or

$$V\dot{H} + \dot{V}H = \Lambda V , \tag{15}$$

Hence

$$(VH)^{\cdot} = \Lambda V . \tag{16}$$

Equations (10) and (13) give

$$(VH_i)^{\cdot} = (VH)^{\cdot} . \tag{17}$$

The last two equations lead to

$$[V(H_i - H)]^{\cdot} = 0 \tag{18}$$

Integration gives

$$H_i = H + \frac{K_i}{V} \ , \ \sum_{i=1}^{3} K_i = 0 \ , \ \sum_{i=1}^{3} K_i^2 = 3K^2 , \tag{19}$$

where $K_i$ are integration constants representing the deviation of the directional Hubble parameters from isotropy, and $K$ is defined in the last quation. Integration of this Equation and use of Equation (3) gives

$$a_i = V^{1/3} \exp\left(K_i \int \frac{1}{V} dt\right). \tag{20}$$

Physicists investigating the Bianchi type I universe models have often defined an average expansion anisotropy by

$$A = \frac{1}{3}\sum_{i=1}^{3}\left(\frac{H_i - H}{H}\right)^2. \tag{21}$$

Using Equation (16), one then obtains

$$K^2 = H^2 V^2 A. \tag{22}$$

Hence, the constant $K$ is a measure of the expansion anisotropy of the universe. The relationship between $K^2$ and $\sigma^2$ is

$$K^2 = \frac{2}{3}V^2\sigma^2. \tag{23}$$

It has lately become more usual to introduce an anisotropy parameter:

$$\Omega_\sigma = \frac{\sigma^2}{3H^2} = \frac{\sigma_0^2}{3H^2 V^2} = \frac{A}{2} \tag{24}$$

As seen from Equation (24), the quantities $\Omega_\sigma$ and $A$ have the same physical interpretation, which is not obvious from the definitions (6) and (21). In this paper, we shall use the parameter $\Omega_\sigma$ and not $A$. However both of these parameters have been much in use and still are, so I have demonstrated their physical identity for the benefit of those who want to read more on the Bianchi type I universe models.

For the present value of $\Omega_\sigma$, I shall use $\Omega_{\sigma 0} \approx 10^{-18}$, which is the upper value permitted by the Planck observations of the temperature fluctuations of the CMB radiation, as suggested by Akarsu et al. [6,7].

Using Equation (6), the anisotropic generalization of Friedmann's first Equation takes the form

$$3H^2 = \kappa\left(\rho_\Lambda + \rho_M + \rho_R + \rho_Z\right) + \frac{\sigma_0^2}{V^2}, \tag{25}$$

where $\rho_\Lambda$, $\rho_M$, $\rho_R$, $\rho_Z$ are the densities of LIVE, cold matter, radiation, and Zeldovich fluid, respectively? A consequence of Einstein's field equations is the laws of energy and momentum conservation in the form

$$T^{\mu\nu}{}_{;\nu} = 0. \tag{26}$$

Assuming that there is no transition between matter and dark energy, this gives the equations of continuity for LIVE, cold matter, radiation, and stiff Zeldovich fluid with the equation of state $p = \rho$,

$$\dot\rho_\Lambda = 0 \;,\; \dot\rho_M + \frac{\dot V}{V}\rho_M = 0 \;,\; \dot\rho_R + \frac{4}{3}\frac{\dot V}{V}\rho_R = 0 \;,\; \dot\rho_Z + 2\frac{\dot V}{V}\rho_Z = 0 \tag{27}$$

Hence

$$\rho_\Lambda = \rho_{\Lambda 0} = \text{constant} \;,\; \rho_M = \rho_{M0}/V \;,\; \rho_R = \rho_{R0}/V^{4/3} \;,\; \rho_Z = \rho_{Z0}/V^2 \tag{28}$$

Where $\rho_{\Lambda 0}$, $\rho_{M0}$, $\rho_{R0}$, and $\rho_{S0}$ are the present values of the average density of LIVE, cold matter, radiation, and Zeldovich fluid.

It has become usual to express Equation (25) in terms of the present values of the density parameters of LIVE, cold matter, radiation, the stiff fluid with $\Omega_{\Lambda 0} = (\kappa/3H_0^2)\rho_{\Lambda 0}$, $\Omega_{M0} = (\kappa/3H_0^2)\rho_{M0}$, $\Omega_{R0} = (\kappa/3H_0^2)\rho_{R0}$, $\Omega_{Z0} = (\kappa/3H_0^2)\rho_{Z0}$, and the anisotropy parameter $\Omega_{\sigma 0} = \sigma_0^2/3H_0^2$. Here, $H_0$ is the present value of the average Hubble parameter, i.e., the Hubble constant of the anisotropic universe model. From Equation (23) and the expression (24) for $\Omega_{\sigma 0}$ we have

$$K^2 = 2H_0^2 \Omega_{\sigma 0}. \tag{29}$$

Inserting these parameters into Equation (25) and putting $t$ equal to the present time $t_0$ gives

$$\Omega_{\Lambda 0} + \Omega_{M0} + \Omega_{R0} + \Omega_{Z0} + \Omega_{\sigma 0} = 1. \tag{30}$$

Inserting Equation (27) into Equation (24) and using that $3H = \dot{V}/V$ gives

$$H^2 = \left[(\Omega_{Z0} + \Omega_{\sigma 0})V^{-2} + \Omega_{R0}V^{-4/3} + \Omega_{M0}V^{-1} + \Omega_{\Lambda 0}\right]H_0^2. \tag{31}$$

Note that the anisotropy appears with the same dependency upon the scale factor as a Zeldovich fluid. Hence, the effect upon the expansion of the universe of the Zeldovich fluid is equivalent to that of the anisotropy of the expansion. In the present article, the Zeldovich fluid will, therefore, be neglected.

If the expansion of the universe is anisotropic, the anisotropy would have a dominating influence on the expansion of the universe very early in the history of the universe.

Saunders was interested in those epochs of the universe that it was possible to observe in the sixties—Rather late in the history of the universe. Then, the radiation term was much less than the matter term in the expression (31). Therefore, he chose to integrate this Equation for an era of the universe where the contribution of radiation could be neglected. He found three solutions: one for $\Lambda > 0$, one for $\Lambda = 0$, and one for $\Lambda < 0$. Since $\Lambda$ represents the density of LIVE we are here only interested in the case $\Lambda > 0$. In this case, Equation (31) reduces to

$$\dot{V} = 3H_0 \left(\Omega_{\Lambda 0}V^2 + \Omega_{M0}V + \Omega_{\sigma 0}\right)^{1/2}. \tag{32}$$

For later comparison, we shall consider four special cases before this Equation is integrated into the general case. We start by presenting the isotropic ΛCDM universe model, following [9], which is presently used as the standard model of the universe.

*2.2. The ΛCDM Universe Model*

Putting $\Omega_{\sigma 0} = 0$ in Equation (32) and integrating gives

$$V_{\Lambda CDM} = \frac{\Omega_{M0}}{\Omega_{\Lambda 0}} \sinh^2 \hat{t} \tag{33}$$

where

$$\hat{t} = \frac{t}{t_\Lambda}, \quad t_\Lambda = \frac{2}{3\sqrt{\Omega_{\Lambda 0}}H_0}. \tag{34}$$

Inserting $\Omega_{\Lambda 0} = 0.3$ and $H_0 = 67.4$ km s$^{-1}$ Mpc$^{-1}$ gives $t_\Lambda \approx 12.5 \cdot 10^9$ years. The age, $t_{\Lambda CDM0}$, of this universe, as defined by normalization of the average scale factor, i.e., of the co-moving volume, $V_{\Lambda CDM}(t_{\Lambda CDM0}) = 1$, is

$$t_{\Lambda CDM0} = t_\Lambda \operatorname{arsinh}\sqrt{\frac{\Omega_{\Lambda 0}}{\Omega_{M0}}} \qquad (35)$$

with $\Omega_{M0} = 0.3$, this gives $t_{\Lambda CDM0} \approx 13.8 \cdot 10^9$ years. It follows from Equations (3), (34), and (35) that the emission point of time of an object with redshift $z$ is

$$t_{\Lambda CDME} = t_* \operatorname{arsinh}\left(\sqrt{\frac{\Omega_{\Lambda 0}}{\Omega_{M0}}} \frac{1}{(1+z)^{3/2}}\right) \quad , \quad t_* = \frac{t_{\Lambda CDM0}}{\operatorname{arsinh}\sqrt{\frac{\Omega_{\Lambda 0}}{\Omega_{M0}}}}. \qquad (36)$$

Inserting $\Omega_{M0} = 0.3$, $\Omega_{\Lambda 0} = 0.7$ and $t_{\Lambda CDM0} = 13.8 \cdot 10^9$ gives $t_* = 11.4 \cdot 10^9$ years and

$$t_{\Lambda CDME} = 11.4 \cdot 10^9 \operatorname{arsinh}\left[\frac{1.53}{(1+z)^{3/2}}\right] \text{years}. \qquad (37)$$

We shall later need the recombination time calculated from this universe model. It corresponds to a redshift $z_{RC} = 1090$ giving $t_{\Lambda CDMRCE} = 4.8 \cdot 10^5$ years. This is not the standard value, i.e., the recombination time when radiation is included in the universe model. It will be calculated below.

Inversely, from Equations (36) and (37), the redshift of radiation emitted at a point of time $t_E$ is

$$z = \left(\frac{\Omega_{\Lambda 0}}{\Omega_{M0}}\right)^{1/3}\left(\frac{1}{\sinh \bar{t}_E}\right)^{2/3} - 1 = \left(\frac{1.53}{\sinh \bar{t}_E}\right)^{2/3} - 1 \quad , \quad \bar{t}_E = \frac{t_E}{t_*}. \qquad (38)$$

It follows from Equations (4) and (33) that the average Hubble parameter is

$$H_{\Lambda CDM} = \sqrt{\Omega_{\Lambda 0}} H_0 \coth \hat{t}. \qquad (39)$$

Using Equations (5) and (39), the deceleration parameter is

$$q_{\Lambda CDM} = \frac{3}{2}\frac{1}{\cosh^2 \hat{t}} - 1 = \frac{1}{2}(1 - 3\tanh^2 \hat{t}). \qquad (40)$$

There is a transition from decelerated expansion to accelerated expansion at the point of time, $t_{\Lambda CDM1}$, defined by $q_{\Lambda CDM}(t_{\Lambda CDM1}) = 0$, leading to

$$\tanh(t_{\Lambda CDM1}) = \frac{1}{\sqrt{3}} \qquad (41)$$

or

$$t_{\Lambda CDM1} = t_\Lambda \operatorname{artanh}\frac{1}{\sqrt{3}}, \qquad (42)$$

giving $t_{\Lambda CDM1} = 7.5 \cdot 10^9$ years. It follows from Equation (42) that

$$\sinh(t_{\Lambda CDM1}) = \frac{1}{\sqrt{2}}. \tag{43}$$

Hence, the corresponding redshift is

$$z(t_{\Lambda CDM1}) = \left(2\frac{\Omega_{\Lambda 0}}{\Omega_{M0}}\right)^{1/3} - 1, \tag{44}$$

giving $z(t_{\Lambda CDM1}) = 0.67$.

According to Equations (25) and (30), the density of the cold matter decreases with time as

$$\rho_M = \frac{\rho_\Lambda}{\sinh^2 \hat{t}}. \tag{45}$$

There is a transition from a matter-dominated era to a LIVE-dominated era at a point of time $t_{\Lambda CDM2}$ given by $\rho(t_2) = \rho_\Lambda$. Hence,

$$t_{\Lambda CDM2} = t_\Lambda \text{arsinh}(1), \tag{46}$$

giving $t_2 = 11 \cdot 10^9$ years. The corresponding redshift is

$$z(t_{\Lambda CDM2}) = \left(\frac{\Omega_{\Lambda 0}}{\Omega_{M0}}\right)^{1/3} - 1, \tag{47}$$

giving $z(t_{\Lambda CDM2}) = 0.33$. Due to the strong repulsive gravitational effect of the negative pressure of LIVE, as seen from the relativistic expression (13) of the effective gravitational mass density, the transition to accelerated cosmic expansion happens before the universe becomes LIVE-dominated.

*2.3. The Kasner Universe*

The *empty anisotropic Bianchi type I universe* has $\Omega_{\Lambda 0} = \Omega_{M0} = 0$ and is called the Kasner universe. For this universe $\Omega_{\sigma 0} = 1$, and Equation (32) reduces to

$$\dot{V}_K = 3H_{K0}. \tag{48}$$

Integrating with $V(0) = 0$ and $V_K(t_{K0}) = 1$ gives

$$V_K = (t/t_{K0}), \tag{49}$$

where $t_{K0} = 1/3H_{K0}$ is the present age of this universe.

Using that

$$\int \frac{1}{V} dt = t_{K0} \ln \frac{t}{t_{K0}}. \tag{50}$$

Equation (20) gives the directional scale factors $a_i$ in the form

$$a_i = \left(\frac{t}{t_{K0}}\right)^{\frac{1}{3} + K_i t_{K0}}. \tag{51}$$

The Hubble parameter of the Kasner universe is

$$H_K = \frac{1}{3t}. \tag{52}$$

*The Hubble horizon* is a surface around an observer separating an internal region where the expansion velocity is less than the velocity of light relative to the observer from an external region where the expansion velocity is larger than that of light. The radius of the Hubble horizon is

$$r_H = \frac{1}{H(t)}, \tag{53}$$

giving

$$r_{HK} = 3t. \tag{54}$$

for the Kasner universe. It follows from Equations (5) and (49) that the deceleration parameter of the Kasner universe is $q_K = 2$, which means cosmic expansion deceleration. This may be somewhat surprising since this universe is empty and the corresponding isotropic universe, the Milne universe, has deceleration equal to zero and hence vanishing cosmic deceleration as expected for an empty universe. Hence, *the anisotropy induces a deceleration of the cosmic expansion*.

In this connection, it may be noted that the isotropic Milne universe is not a special case of the Bianchi type I universe models because these models have vanishing spatial curvature, while the Milne universe has negative spatial curvature.

*2.4. The LIVE-Dominated Bianchi Type I Universe*

Let us first consider the LIVE-dominated *isotropic* case with $\Omega_{\sigma 0} = \Omega_{M0} = 0$. This is the *De Sitter universe*. Then, Equation (32) reduces to

$$\dot{V}_{DS} = 3H_{DS0}\sqrt{\Omega_{\Lambda 0}}V_{DS}. \tag{55}$$

This universe model does not permit the initial condition $V(0) = 0$. Integration with the boundary condition $V(t_0) = 1$ gives the co-moving volume

$$V_{DS} = e^{3H_{DS0}(t-t_0)}, \tag{56}$$

and constant Hubble parameter $H = H_{DS0}$.

Next we shall consider the *anisotropic* LIVE-dominated Bianchi type I universe. This universe model is particularly relevant as a model of the early part of the inflationary era since the relationship (7) implies that even a very small anisotropy at the present time means that the anisotropy may have been great, and even dominating, during the first part of the inflationary era. It was thoroughly described in ref. [9], and here I shall only recapitulate the main properties of this model.

In this model $\Omega_{M0} = 0$. Then, the solution of Equation (32) takes the form

$$V_{BL} = \sqrt{\frac{\Omega_{\sigma 0}}{\Omega_{\Lambda 0}}}\sinh(2\hat{t}). \tag{57}$$

It follows from Equations (4) and (56) that in the LIVE-dominated universe, the Hubble parameter is

$$H_{BL} = \sqrt{\Omega_{\Lambda 0}}H_{ISO0}\coth(2\hat{t}), \tag{58}$$

which is similar to the corresponding formula (39) for the Hubble parameter in the $\Lambda$CDM universe, but with $\coth(2\hat{t})$ instead of $\coth\hat{t}$.

In the calculation of the directional scale factors, $a_i$, from Equation (20), we need the integral

$$\int \frac{1}{V} dt = \frac{1}{3\sqrt{\Omega_{\sigma 0}} H_0} \ln(\tanh\hat{t}) \tag{59}$$

The resulting expression for $a_i$ can be simplified by introducing the constants

$$p_i = \frac{1}{3}\left(1 + \frac{K_i}{\sqrt{\Omega_{\sigma 0}} H_0}\right) \tag{60}$$

Using that $\sum_{i=0}^{3} K_i = 0$, $\Omega_{\sigma 0} H_0^2 = \sigma_0^3/3$ and Equation (13) with $V_0 = 1$ we find that the constants $p_i$ fulfil the relationships

$$\sum_{i=1}^{3} p_i = 1 \quad , \quad \sum_{i=1}^{3} p_i^2 = 1 \tag{61}$$

This leads to the following expression for the directional scale factors:

$$a_i = \left(4\frac{\Omega_{\sigma 0}}{\Omega_{\Lambda 0}}\right)^{1/6} \sinh^{p_i}\hat{t} \cosh^{\frac{2}{3}-p_i}\hat{t} \,. \tag{62}$$

In agreement with Equation (22) in ref. [9], this LIVE-dominated Bianchi type I universe model has been generalized to include viscosity by Mostafapoor and Grøn [12].

The relationships (61) mean that $p_1$ and $p_2$ can be expressed in terms of $p_3$ as follows:

$$p_1 = \frac{1}{2}\left[1 - p_3 + \sqrt{(3p_3+1)(1-p_3)}\right] \quad , \quad p_2 = \frac{1}{2}\left[1 - p_3 - \sqrt{(3p_3+1)(1-p_3)}\right] \tag{63}$$

There are two cases, $p_3 = -1/3$, $p_1 = p_2 = 2/3$ and $p_3 = 1$, $p_1 = p_2 = 0$ permitting two equal scale factors in this very early LIVE and anisotropy-dominated era. The first one has scale factors

$$a_1 = a_2 = \left(4\frac{\Omega_{\sigma 0}}{\Omega_{\Lambda 0}}\right)^{1/6} \sinh^{2/3}\hat{t} \quad , \quad a_3 = \left(4\frac{\Omega_{\sigma 0}}{\Omega_{\Lambda 0}}\right)^{1/6} \frac{\cosh\hat{t}}{\sinh^{1/3}\hat{t}} \,. \tag{64}$$

The second model has scale factors

$$a_1 = a_2 = \left(4\frac{\Omega_{\sigma 0}}{\Omega_{\Lambda 0}}\right)^{1/6} \cosh^{2/3}\hat{t} \quad , \quad a_3 = \left(4\frac{\Omega_{\sigma 0}}{\Omega_{\Lambda 0}}\right)^{1/6} \frac{\sinh\hat{t}}{\cosh^{1/3}\hat{t}} \tag{65}$$

The models initially behave rather differently. Two scale factors of the first model initially vanish, and the third is infinite. Hence, this model starts as a needle singularity and approaches isotropy at late times. One scale factor of the second model initially vanishes, and two are infinite. Thus, this model begins as a plane singularity with vanishing thickness and with infinitely great extension. But like the first model it approaches isotropy at late times. This indicates that very different initial behavior at the beginning of the inflationary era would not lead to corresponding late time differences in the models. The anisotropic models evolve in an exponential way towards isotropy.

It follows from Equations (24), (57), and (58) that in this universe model, the anisotropy parameter varies with time as

$$\Omega_\sigma = \frac{1}{\cosh^2(2\hat{t})}. \tag{66}$$

Hence, initially, the universe had the maximal value $\Omega_\sigma(0) = 1$ equal to that of the empty Kasner universe. The time of recombination, $t_{ISORC} = 380000$ years or $\hat{t}_{ISORC} = 3.0 \cdot 10^{-5}$ was determined using the isotropic ΛCDM universe model. It will be shown below that the upper bound on the anisotropy of the cosmic expansion is so small that the effect of the anisotropy upon the calculated value of the recombination time is negligible. Hence, we shall use the value above in the main part of this article. At the time of the recombination the anisotropy parameter was still $\Omega_\sigma(t_{ISORC}) \approx 1$. For the present time, $\hat{t}_0 = t_0/t_\Lambda = 1.104$, this formula predicts $\Omega_\sigma(t_0) = 0.047$ which is much larger than allowed by the observational restriction found by Akarsü et al. [7]. We shall later see how the inclusion of mass in the universe model modifies this result.

In the early era with $t \ll t_\Lambda$, we can make the approximations $\sinh(2\hat{t}) \approx 2\hat{t}$, $\coth(2\hat{t}) \approx 1/2\hat{t}$ in Equations (57) and (58). This gives

$$V_{BL} \approx 3\sqrt{\Omega_{\sigma 0}}t \quad , \quad H_{BL} \approx \frac{1}{3t} \tag{67}$$

in agreement with Equations (49) and (52) with $\Omega_{\sigma 0} = 1$. At late times, $t \gg t_\Lambda$, the anisotropy decreases exponentially, and the universe model approaches the de Sitter universe.

*2.5. The Anisotropic Generalization of the ΛCDM Universe Model*

We now go back to the general case with cold matter, LIVE, and anisotropy. Introducing a new variable,

$$y = \frac{2\Omega_{\Lambda 0}}{\sqrt{\Omega_{M0}^2 - 4\Omega_{\Lambda 0}\Omega_{\sigma 0}}}\left(V + \frac{1}{2}\frac{\Omega_{M0}}{\Omega_{\Lambda 0}}\right), \tag{68}$$

Equation (32) takes the form

$$\frac{dy}{\sqrt{y^2 - 1}} = 3\sqrt{\Omega_{\Lambda 0}}H_0 dt \tag{69}$$

Integrating this Equation with the initial condition $V(0) = 0$ gives

$$V = A\cosh(2\hat{t} + C) - B \quad ,$$
$$A = B\sqrt{1 - 4K_\sigma^2} \quad , \quad B = \frac{1}{2}\frac{\Omega_{M0}}{\Omega_{\Lambda 0}} \quad , \quad K_\sigma = \frac{\sqrt{\Omega_{\Lambda 0}\Omega_{\sigma 0}}}{\Omega_{M0}} \quad , \quad \cosh C = \frac{B}{A} \quad , \tag{70}$$

where $\hat{t}$ is given in equation (34), and the cosmological constant is $\Lambda = \kappa\rho_{\Lambda 0} = 3\Omega_{\Lambda 0}H_0^2$. Expression (70) shows that for $\hat{t} \gg 1$, this universe goes into an era with eternal exponential expansion.

The present values of the mass parameters of matter and dark energy are $\Omega_{\Lambda 0} = 0.7$ and $\Omega_{M0} = 0.3$. From the baryonic acoustic oscillations and cosmic microwave background (CMB) data, Akarsu and co-workers [6,7] obtained the constraint $\Omega_{\sigma 0} <$ $10^{-18}$. Furthermore, they wrote: "Demanding that the expansion anisotropy has no significant effect on the standard big bang nucleosynthesis (BBN), we find the constraint $\Omega_{\sigma 0} \leq 10^{-23}$." I shall here use $\Omega_{\sigma 0} \approx 10^{-18}$.

The value of the Hubble constant, as given by the Planck project, is $H_0 = (67.4 \pm 0.5)$ km s$^{-1}$ Mpc$^{-1}$. This gives $A \approx B \approx 0.21$, $K_\sigma = 2.8 \cdot 10^{-9}$, $C = 1.1 \cdot 10^{-7}$, $t_\Lambda \approx 11 \cdot 10^9$ years. Using that the age of the universe is $t_0 = 13.8 \cdot 10^9$ years, the present value of $\hat{t}$ is $\hat{t}_0 = 1.25$. The universe became transparent to the CMB background radiation at a point of time $t_1 = 3.8 \cdot 10^5$ years, corresponding to $\hat{t}_1 = 3.4 \cdot 10^{-5}$.

Equation (70) may also be written as

$$V = \frac{\Omega_{M0}}{\Omega_{\Lambda 0}} \sinh^2 \hat{t} + \sqrt{\frac{\Omega_{\sigma 0}}{\Omega_{\Lambda 0}}} \sinh(2\hat{t}). \tag{71}$$

This is the most useful expression for the time-dependence of the co-moving volume in the Bianchi type I anisotropic generalization of the ΛCDM-universe model since it separates nicely the effects of anisotropy and matter upon the time evolution of the co-moving volume. Formula (71) shows that the co-moving volume of the anisotropic generalization of the ΛCDM-model is the sum of the co-moving volume of the isotropic ΛCDM-model as given in Equation (33) and the anisotropic LIVE-dominated Bianchi type I universe as given in Equation (57). Expression (71) shows that the universe comes from an initial singularity with $\lim_{t \to 0} V = 0$.

Differentiating Equation (71) and using Equation (4) gives the average Hubble parameter of the anisotropic generalization of the ΛCDM universe as

$$H = \frac{1}{2}\sqrt{\Omega_{\Lambda 0}} H_0 \frac{\sinh(2\hat{t}) + 2K_\sigma \cosh(2\hat{t})}{\sinh^2 \hat{t} + K_\sigma \sinh(2\hat{t})} \tag{72}$$

This expression can be written as

$$H = \sqrt{\Omega_{\Lambda 0}} H_0 \frac{1 + 2K_\sigma \coth(2\hat{t})}{\tanh \hat{t} + 2K_\sigma}. \tag{73}$$

These expressions, together with the expression for $\hat{t}$ in Equation (70), show that the universe started from a Kasner-like era with Hubble parameter as given in Equation (52).

The average deceleration parameter, $q$, of this universe model is

$$q = 6 \frac{\sinh^2 \hat{t} + K_\sigma \sinh(2\hat{t}) + 2K_\sigma^2}{\left[\sinh(2\hat{t}) + 2K_\sigma \cosh(2\hat{t})\right]^2} - 1, \tag{74}$$

In the early era with $\hat{t} \ll 1$, this expression approaches that of the Kasner era, $q_K = 2$.

From Equations (7) and (71), the time-evolution of the shear scalar is

$$\sigma^2 = \frac{\sigma_0^2}{\left[\frac{\Omega_{M0}}{\Omega_{\Lambda 0}} \sinh^2 \hat{t} + \sqrt{\frac{\Omega_{\sigma 0}}{\Omega_{\Lambda 0}}} \sinh(2\hat{t})\right]^2}. \tag{75}$$

This shows that for $\hat{t} \gg 1$, the shear scalar decreases exponentially with time. Using Equations (71) and (72), expression (24) for the anisotropy parameter can be written as

$$\Omega_\sigma = \frac{1}{\left[\cosh(2\hat{t}) + (1/2K_\sigma)\sinh(2\hat{t})\right]^2}. \tag{76}$$

Again, we see that the initial value of the anisotropy parameter is $\Omega_\sigma(0) = 1$. The two terms in the denominator are equal at a point of time

$$t_4 = (t_\Lambda/2)\operatorname{artanh}(2K_\sigma). \tag{77}$$

Inserting $t_\Lambda = 12.5 \cdot 10^9$ years and $K_\sigma = 2.8 \cdot 10^{-9}$ gives $t_4 = 38$ years. At this point of time, the anisotropy parameter has the value

$$\Omega_\sigma(t_4) = 0.25 - K_\sigma^2 \approx 0.25. \tag{78}$$

From then on the anisotropy decreases at an increasing rate, and the value of the anisotropy parameter at the point of time of the recombination, $t_{RC} = 380,000$ years or $\hat{t}_{RC} = 3.3 \cdot 10^{-5}$, is $\Omega_\sigma(t_{RC}) = 8.3 \cdot 10^{-9}$. The value at the present time, $\hat{t}_0 = 1.104$, as given in Equation (76), is $\Omega_\sigma(t_0) = 1.5 \cdot 10^{-18}$.

We proceed to calculate the directional scale factors in the same way as above for the LIVE-dominated Bianchi type I universe. We then need the integral

$$\int \frac{1}{V} dt = -\frac{1}{3\sqrt{\Omega_{\sigma 0}} H_0} \ln\left(\frac{\Omega_{M0}}{\Omega_{\sigma 0}} + 2\sqrt{\frac{\Omega_{\sigma 0}}{\Omega_{\Lambda 0}}} \coth \hat{t}\right). \tag{79}$$

Again, the resulting expression for $a_i$ can be simplified by introducing the constants in Equation (60), fulfilling the relationships (61). From Equations (20), (71), and (79), we obtain

$$a_i = \left(\frac{\Omega_{M0}}{\Omega_{\Lambda 0}} \sinh \hat{t} + 2\sqrt{\frac{\Omega_{\sigma 0}}{\Omega_{\Lambda 0}}} \cosh \hat{t}\right)^{p_i} \sinh^{\frac{2}{3} - p_i} \hat{t}. \tag{80}$$

Let us here, too, consider the two cases permitted by Equation (63) with two equal scale factors. The first one is $p_3 = -1/3$ giving $p_1 = p_2 = 2/3$. In this case, the directional scale factors are

$$a_1 = a_2 = \left(\frac{\Omega_{M0}}{\Omega_{\Lambda 0}} \sinh \hat{t} + 2\sqrt{\frac{\Omega_{\sigma 0}}{\Omega_{\Lambda 0}}} \cosh \hat{t}\right)^{2/3}, \quad a_3 = \frac{\sinh \hat{t}}{\left(\frac{\Omega_{M0}}{\Omega_{\Lambda 0}} \sinh \hat{t} + 2\sqrt{\frac{\Omega_{\sigma 0}}{\Omega_{\Lambda 0}}} \cosh \hat{t}\right)^{1/3}} \tag{81}$$

The evolution of this model can be separated into three periods. There is a transition between the first two periods at a point of time $t_1$ when $a_1(t_1) = a_3(t_1)$. This gives

$$t_1 = t_\Lambda \operatorname{artanh} \frac{2\sqrt{\Omega_{\Lambda 0} \Omega_{\sigma 0}}}{\Omega_{\Lambda 0} - \Omega_{M0}}. \tag{82}$$

Inserting $\Omega_{\Lambda 0}=0.7$, $\Omega_{M0}=0.3$, $\Omega_{\sigma 0}\simeq 4\cdot 10^{-14}$ and $t_{\Lambda}=11\cdot 10^9$ years gives $t_1=9240$ years. At this point of time, the universe is isotropic. The universe starts from a plane singularity with $a_3=0$ and has a plane-like character for $t<t_1$. For $t>t_1$ and until $t$ approaches $t_{\Lambda}$, the universe has $a_1=a_2<a_3$. Then, it has a needle-like character. For $t>t_{\Lambda}$, the universe approaches isotropy exponentially.

The other case with two equal scale factors, has $p_3=1$, giving $p_1=p_2=0$. Then, the directional scale factors are

$$a_1=a_2=\sinh^{2/3}\hat{t} \quad , \quad a_3=\left(\frac{\Omega_{M0}}{\Omega_{\Lambda 0}}\sinh\hat{t}+2\sqrt{\frac{\Omega_{\sigma 0}}{\Omega_{\Lambda 0}}}\cosh\hat{t}\right)\sinh^{-\frac{1}{3}}\hat{t}. \tag{83}$$

The transition time between the two first periods is the same as in the first case, but the behavior is the opposite. The universe starts from a needle singularity; there is a transition at $t=t_1$ through an isotropic moment to a plane-like era, and for $t>t_{\Lambda}$, the universe approaches isotropy exponentially.

Also, in the general case with different scale factors in all directions, the evolution of this universe model can be separated into three periods with different behaviors.

1. *Anisotropy dominated era*. Expression (76) for the anisotropy parameter shows that the universe started from a state with maximal anisotropy, $\lim_{t\to 0}\Omega_{\sigma}=1$, equal to the constant anisotropy of an empty Kasner universe with vanishing cosmological constant, and ends in an isotropic state with $\Omega_{\sigma}=0$. The initial value of the deceleration parameter was $q(0)=2$, which means a large deceleration. Hence, the universe must have started by a process not described by this solution, which has given the universe an initial expansion velocity. At the end of this singular process, which marks the beginning of the evolution described by the anisotropic ΛCDM universe model, the Hubble parameter had an infinitely large value. The universe then entered an anisotropy-dominated era with Kasner-like behavior. In this era $\hat{t}\ll 1$ and we can use the approximations $\sinh\hat{t}\approx\hat{t}$, $\cosh\hat{t}\approx 1$. To first order in $\hat{t}$ Equations (71) and (73) then give

$$V\approx 2\sqrt{\frac{\Omega_{\sigma 0}}{\Omega_{\Lambda 0}}}\hat{t}=3\sqrt{\Omega_{\Lambda 0}}H_0 t \quad , \quad H\approx\frac{1}{2}\sqrt{\Omega_{\Lambda 0}}H_0\frac{1}{\hat{t}}=\frac{1}{3t} \tag{84}$$

in agreement with Equations (49) and (52);

2. *Matter dominated era*. The transition to a matter-dominated era happened at a time $t_2$ when the two terms of the expression (71) for the co-moving volume had the same size, i.e.,

$$t_2=t_{\Lambda}\operatorname{artanh}(2K_{\sigma}), \tag{85}$$

giving $t_2=1.4\cdot 10^4$ years. The values of the co-moving volume, Hubble parameter, and the anisotropy parameter at this point of time was

$$V(t_2)\approx 8\frac{\Omega_{\sigma 0}}{\Omega_{M0}} \quad , \quad H(t_2)\approx\frac{3}{8}\frac{\Omega_{M0}}{\sqrt{\Omega_{\sigma 0}}}H_0 \quad , \quad q(t_2)\approx 1.67, \tag{86}$$

giving $V(t_2)=1.1\cdot 10^{-12}$ and $H(t_2)=5.6\cdot 10^5 H_0$. The redshift of radiation emitted at the time $t_2$ is $z(t_2)\approx 10^4$.

The value of $\Omega_\sigma$ at the recombination time $t_{RC} = 3.8 \cdot 10^5$ years, i.e., $\hat{t}_{RC} = 3.3 \cdot 10^{-5}$, when the universe became transparent for the CMB background radiation, as given by Equation (76), was $\Omega_\sigma(t_{RC}) = 1.8 \cdot 10^{-6}$. This happened quite early in the matter-dominated era;

3. *LIVE dominated era.* The transition from a matter-dominated to a LIVE-dominated era is here defined by the condition that cosmic deceleration due to the attractive gravity of the matter changes to accelerated expansion due to the repulsive gravity of the LIVE. Hence, the point of time, $t_3$, of this transition is given by the condition that the deceleration parameter vanishes, $q(t_3) = 0$.

In the case of the isotropic universe, this transition happens at a point of time, $t_{\Lambda CDM2}$, given by Equation (46) leading to $t_{\Lambda CDM2} = 7.5 \cdot 10^9$ years. Since $\hat{t}_3$ is rather large, $\hat{t}_3 \approx 0.6$, and $K_\sigma$ is very small, we can calculate the influence of the expansion anisotropy upon the value of $t_3$ with good accuracy by linearizing expression (62) in $K_\sigma$. This gives

$$q \approx q_{\Lambda CDM} - \frac{3K_\sigma}{\sinh \hat{t} \cosh^3 \hat{t}}. \tag{87}$$

where $q_{\Lambda CDM}$ is given in Equation (40). It is shown in Appendix A that this condition leads to the result that the change of the transition time due to the expansion anisotropy is not greater than $\Delta t_3 = 1.4 \cdot 10^{-6} t_\Lambda = 1.5 \cdot 10^4$ years. Hence the transition time is close to that of the corresponding isotropic universe given in Equation (46). The corresponding redshift is

$$z(t_3) = V^{-1/3}(t_3) - 1 \approx \left(2\frac{\Omega_{\Lambda 0}}{\Omega_{M0}}\right)^{1/3}\left(1 - \frac{2}{\sqrt{3}}K_\sigma\right) - 1. \tag{88}$$

Inserting the values of the constants gives $z(t_3) \approx 0.67$ like the transition redshift in the isotropic universe.

In the anisotropic universe the age of the universe is given by applying the normalization condition $V(t_0) = 1$ to $V$ in Equation (71). This gives

$$\sinh^2 \hat{t}_0 = \frac{\Omega_{\Lambda 0}\left(\Omega_{M0} + 2\sqrt{\Omega_{\sigma 0}} + 2\Omega_{\sigma 0}\right)}{\Omega_{M0}^2 - 4\Omega_{\Lambda 0}\Omega_{\sigma 0}}. \tag{89}$$

To 1. order in $\sqrt{\Omega_{\sigma 0}}$ this gives

$$\sinh^2 \hat{t}_0 = \frac{\Omega_{\Lambda 0}}{\Omega_{M0}}\left(1 + 2\frac{\sqrt{\Omega_{\sigma 0}}}{\Omega_{M0}}\right). \tag{90}$$

In the case of the isotropic ΛCDM-universe, this reduces to

$$\sinh^2 \hat{t}_{\Lambda CDM0} = \frac{\Omega_{\Lambda 0}}{\Omega_{M0}}. \tag{91}$$

In accordance with the standard expression, (35), for the age of the isotropic ΛCDM universe model.

It follows from Equations (90) and (91) that

$$t_{\Lambda CDM0} - t_0 \approx \left(\frac{\Omega_{\sigma 0}}{\Omega_{M0}^3}\right)^{1/2} t_\Lambda \quad (92)$$

Inserting the values for the constants as given above we obtain $t_0 - t_{\Lambda CDM0} \approx 3.5 \cdot 10^4$ years.

**3. Review of Some Papers on the Bianchi Type I Universe Models**

In this Section I shall review some selected articles discussing Bianchi type I universe models in light of the results in the previous sections.

*3.1. Anisotropic Brane Universe Models*

C. M. Chen, T. Harko, and M. K. Mak [13] have discussed anisotropic brane universe models. They obtained solutions of Einstein's field equations describing several types of Universe models. I shall here focus on the model most closely related to the present work. They argued that the equation of state most appropriate to describe the high-density regime of the early Universe is the stiff Zeldovich one, with $p = \rho$. In the case of ordinary general theory of relativity with no brane their constant $k_5 = 0$. Then their solutions of the field equations reduce to those in Section 2.4 above.

*3.2. Anisotropic Universe Models with Scalar Field and Phantom Field*

In 2008 B. C. Paul and D. Paul published a paper [14] where they investigated some Bianchi type I universe models with non-vanishing cosmological constant and either a scalar field or a phantom field. In the case where these fields can be neglected compared to the LIVE represented by a cosmological constant, their model reduces to the one considered in [9].

On 8 January 2024, Mark P. Herzberg and Abraham Loeb published a preprint [15] with the title *Constraints on an Anisotropic Universe*. They first considered a matter dominated Bianchi type I universe with vanishing cosmological constant. Then, Equation (32) reduces to

$$\dot{V} = \left(\Omega_{M0} V + \Omega_{\sigma 0}\right)^{1/2} 3H_{M0} \quad (93)$$

Integration with $V(0) = 0$ gives

$$V(t) = \left(\frac{3}{4}\Omega_{M0} H_{M0} t + \sqrt{\Omega_{\sigma 0}}\right) 3\Omega_{M0} H_{M0} t \quad (94)$$

This Equation shows that there is a transition from an early period where the anisotropy dominates to a late period where matter dominates at the point of time

$$t_{A \to M} = \frac{4\sqrt{\Omega_{\sigma 0}}}{3\Omega_{M0}} \frac{1}{H_{M0}} \quad (95)$$

Inserting $1/H_{M0} = 1.4 \cdot 10^{10}$ years, $\Omega_{\sigma 0} \approx 10^{-18}$ and $\Omega_{M0} \approx 0.3$ gives $t_{A \to M} \approx 56$ years. Equations (28) and (94) show that in the early anisotropy-dominated era, the matter density decreases as $1/t$, and in the later matter-dominated era, the density decreases as $1/t^2$, as it does in the isotropic universe.

Inserting Equation (94) into Equation (20) and introducing the constants

$$p_i = \frac{1}{3}\left(1 + \frac{K_i}{H_0 t_0} \frac{\Omega_{M0}}{\sqrt{\Omega_{\sigma 0}}}\right), \quad (96)$$

fulfilling Equation (49) leads to

$$a_i = 4^{\frac{1}{3}}\left(\frac{3}{4}\Omega_{M0}H_{M0}t + \sqrt{\Omega_{\sigma 0}}\right)^{p_i}\left(\frac{3}{4}\Omega_{M0}H_{M0}t\right)^{\frac{2}{3}-p_i} \tag{97}$$

Herzberg and Abraham Loeb considered a model with two equal scale factors. Again, Equation (63) shows that there are two possibilities with two equal scale factors: The first one is $p_3 = -1/3$, giving $p_1 = p_2 = 2/3$. In this case, the directional scale factors are

$$a_1 = a_2 = \left(\frac{3}{2}\Omega_{M0}H_{M\|0}t + 2\sqrt{\Omega_{\sigma 0}}\right)^{2/3} \quad , \quad a_3 = \frac{(3/2)\Omega_{M0}H_{M\|0}t}{\left(\frac{3}{2}H_{M\|0}\Omega_{M0}t + 2\sqrt{\Omega_{\sigma 0}}\right)^{1/3}} \tag{98}$$

where $H_{M\|0}$ is the Hubble constant with two equal scale factors. Although the initial co-moving volume vanishes, the scale factors $a_1$ and $a_2$ are non-vanishing at the initial moment and then form a plane while $a_3(0) = 0$. Hence, this universe model starts from an initial plane-like singularity. There is no transition to a needle-like period with $a_3 > a_1$ since $a_1 = a_2 > a_3$ for all values of $t$.

The other case with two equal scale factors is $p_3 = 1$ giving $p_1 = p_2 = 0$. Then, the directional scale factors are

$$a_1 = a_2 = 4^{1/3}\left(\frac{3}{4}\Omega_{M0}H_{M\|0}t\right)^{2/3} \quad , \quad a_3 = 4^{1/3}\left(\frac{3}{4}\Omega_{M0}H_{M\|0}t + \sqrt{\Omega_{\sigma 0}}\right)\left(\frac{3}{4}\Omega_{M0}H_{M\|0}t\right)^{-1/3} \tag{99}$$

In this model, $\lim_{t\to 0}a_1 = \lim_{t\to 0}a_2 = 0$, $\lim_{t\to 0}a_3 = \infty$. Hence, there is an initial singularity with a needle-like character. Similarly, as in the first case, there is no transition to a plane-like period since $a_1 = a_2 < a_3$ for all values of $t$.

In the isotropic case with $\Omega_{\sigma 0} = 0$ this mass-dominated Bianchi type I universe model reduces to the Einstein de Sitter-universe with mass parameter of the cold matter $\Omega_{M0} = 1$, and the formulae for the co-moving volume and the scale factor reduce to

$$V(t) = \left(\frac{3}{2}H_{MIS00}t\right)^2 \quad , \quad a(t) = \left(\frac{3}{2}H_{MIS00}t\right)^{2/3} \tag{100}$$

The age of this universe model is found either from $V(t_0) = 1$ or by calculating the present value of the Hubble parameter and is

$$t_{MIS00} = \frac{2}{3}\frac{1}{H_{MIS00}} \tag{101}$$

Similarly, one finds the age of the corresponding anisotropic universe model by inserting $V(t_0) = 1$ in Equation (94) and using that $\Omega_{M0} = 1 - \Omega_{\sigma 0}$, giving

$$t_{M0} = \frac{\sqrt{1+\Omega_{\sigma 0}} - \sqrt{\Omega_{\sigma 0}}}{1-\Omega_{\sigma 0}}t_{MISO} \tag{102}$$

To 1. order in $\sqrt{\Omega_{\sigma 0}}$ this gives

$$t_{M0} \approx \left(1 - \sqrt{\Omega_{\sigma 0}}\right) t_{MISO0} \tag{103}$$

Hence, the anisotropy of the expansion reduces the age of the universe by

$$t_{MISO0} - t_{M0} \approx \sqrt{\Omega_{\sigma 0}} t_{MISO} \tag{104}$$

Akarsu et al. [6,7] have found that $\Omega_{\sigma 0} < 4 \cdot 10^{-18}$. Hence, with $t_{MISO0} = 1.4 \cdot 10^{10}$ years, the anisotropy decreases the age by less than 28 years.

The average Hubble parameter of the matter-dominated universe model with two equal scale factors is found by differentiation of Equation (94):

$$H_{M\parallel} = \frac{1}{3}\frac{\dot{V}}{V} = \frac{2}{3}\frac{\Omega_{M0}H_{M\parallel 0}t + (2/3)\sqrt{\Omega_{\sigma 0}}}{\Omega_{M0}H_{M\parallel 0}t + (4/3)\sqrt{\Omega_{\sigma 0}}}\frac{1}{t} \tag{105}$$

The directional Hubble parameters of the anisotropic, mass-dominated universe model are

$$H_1 = H_2 = \frac{2}{3}\frac{\Omega_{M0}}{\Omega_{M0}H_{M\parallel 0}t + (4/3)\sqrt{\Omega_{\sigma 0}}}H_{M\parallel 0} ,$$

$$H_3 = \frac{2}{3}\frac{\Omega_{M0}H_{M\parallel 0}t + 2\sqrt{\Omega_{\sigma 0}}}{\Omega_{M0}H_{M\parallel 0}t + (4/3)\sqrt{\Omega_{\sigma 0}}}\frac{1}{t} \tag{106}$$

It may be noted that $H_3$ can be written as

$$H_3 = \left(1 + \frac{2\sqrt{\Omega_{\sigma 0}}}{\Omega_{M0}H_{M\parallel 0}t}\right) H_1 \tag{107}$$

Hence, the difference between $H_1$ and $H_3$ decreases as $1/t$ in agreement with a result found by Herzberg and Loeb [15].

In the fully asymmetric case, they write: "Now let us consider the more general case in which all the scale factors are different. In this case, we do not have an analytical solution to the above equations since the equations are all coupled. Nevertheless, we can solve the equations numerically." The reason for this is an unfortunate way of writing the field equations. As shown above, there are analytic solutions in this general case, too, for a matter-dominated universe, both with and without a cosmological constant. These were originally found by Saunders [8], although in another form than in the present paper.

Herzberg and Loeb [15] also considered a radiation-dominated universe with $\Omega_{\Lambda 0} = \Omega_{M0} = \Omega_{\sigma 0} = 0$. Then, Equation (32) reduces to

$$a^2\dot{a} = \sqrt{\Omega_{R0}a^2 + \Omega_{\sigma 0}}H_0 \tag{108}$$

The solution of this Equation with $a(0) = 0$ is

$$a\sqrt{\Omega_{R0}a^2 + \Omega_{\sigma 0}} - \frac{\Omega_{\sigma 0}}{\sqrt{\Omega_{R0}}}\operatorname{arsinh}\left(\sqrt{\frac{\Omega_{R0}}{\Omega_{\sigma 0}}}a\right) = 2\Omega_{R0}H_0 t \tag{109}$$

The age of this universe is given by $a(t_{R0}) = 1$. From this together with $\Omega_{R0} + \Omega_{\sigma 0} = 1$ we obtain

$$t_{R0} = \left(1 - \frac{\Omega_{\sigma 0}}{\sqrt{\Omega_{R0}}} \operatorname{arsinh} \sqrt{\frac{\Omega_{R0}}{\Omega_{\sigma 0}}}\right) \frac{t_{RIS00}}{\Omega_{R0}} \quad (110)$$

where

$$t_{RIS00} = \frac{1}{2H_0} \quad (111)$$

is the age of the corresponding radiation-dominated isotropic universe. Hence, the expansion anisotropy increases the age a little. With $t_{RIS00} = 13.8 \cdot 10^9$ years, $\Omega_{R0} = 5 \cdot 10^{-4}$, $\Omega_{\sigma 0} \approx 10^{-18}$ Equation (98) gives $t_{RIS00} - t_{R0} \approx$ a few minutes.

It should be noted, however, that this model cannot give a realistic description of the universe. It requires that $\Omega_{\sigma 0} = 1 - \Omega_{R0}$. This means that the observed value $\Omega_{R0} = 5 \cdot 10^{-4}$ requires $\Omega_{\sigma 0} \approx 1$ which is 23 orders of magnitude greater than that permitted by the effect of the expansion anisotropy upon the cosmic nucleosynthesis.

**4. A Simple Bianchi Type I Universe Applied to the Hubble Tension**

In this Section, we shall first consider the most simple universe model of this type, where only one direction expands differently from the others, following M. Le Delliou et al. [16].

A model with a general perfect fluid, including pressure, was considered. However, it should be noted that the pressure of the radiation appeared only in the Einstein Equations (3)–(5) in ref. [7], not in the integrated equations. Hence, putting $p = 0$, i.e., neglecting pressure, makes no changes in their results.

The line element has the form (using units so that the velocity of light $c = 1$)

$$ds^2 = -dt^2 + a^2(t)\left[(1 + \varepsilon(t))^2 dx^2 + dy^2 + dz^2\right], \quad (112)$$

where the scale factor $a(t)$ is normalized so that its present value is $a(t_0) = 1$. The departure from isotropy is measured by the anisotropic perturbation parameter $\varepsilon$. In accordance with observational constraints, Delliou et al. applied the initial condition $\varepsilon_{re} = 10^{-5}$ at the point of time of the cosmic recombination, $t_{re} = 380,000$ years, when the scale factor had the value $a_{re} \simeq 10^{-3}$. It is shown below that $\varepsilon$ is a decreasing function of time, so its present value is $\varepsilon_0 < 10^{-5}$.

The authors have given the following formula for the evolution of the average Hubble parameter of this anisotropic universe model in terms of the scale factor and the anisotropy parameter and its rate of change,

$$H = H_{ISO0}\sqrt{\Omega_{M0}\left(\frac{1}{a^3}\frac{1+\varepsilon_0}{1+\varepsilon} - 1\right) + 1 + \frac{2}{3H_{ISO0}}\frac{\dot{\varepsilon}_0}{1+\varepsilon_0}}. \quad (113)$$

Hence, the Hubble constant of this anisotropic universe model is

$$H_0 = H_{ISO0}\sqrt{1 + \frac{2}{3H_{ISO0}}\frac{\dot{\varepsilon}_0}{1+\varepsilon_0}}. \quad (114)$$

The last term is much less than 1, so we can, with sufficient accuracy, use a series expansion to first order in this term, giving

$$H_0 \approx H_{ISO0} + \frac{1}{3}\frac{\dot{\varepsilon}_0}{1+\varepsilon_0} \quad (115)$$

Thus, the difference between the Hubble constant in this universe model with anisotropic expansion in one direction and in a universe with isotropic expansion is with good accuracy

$$\Delta H_0 = (1/3)\dot{\varepsilon}_0 \quad (116)$$

The present value of $\dot{\varepsilon}_0$ coming from observations can be estimated from the formulae in the appendix of [16]. According to their Equation (A8),

$$\dot{\varepsilon}_0 = \frac{\varepsilon_0}{\Delta_0} H_0 \quad (117)$$

with

$$\Delta_0 = \frac{2}{\Omega_{M0}}\sqrt{\Omega_{M0}\left(a_{re}^{-3}-1\right)+1} - 1 \quad (118)$$

Since $a_{re} \approx 10^{-3}$, we can approximate this expression with

$$\Delta_0 \approx \frac{2}{\sqrt{\Omega_{M0}}} a_{re}^{-3/2} \quad (119)$$

Hence, since we are only interested in an order of magnitude estimate, we can calculate the value of the present rate of change of the anisotropy parameter from

$$\dot{\varepsilon}_0 \approx (1/2)\sqrt{\Omega_{M0}} a_{re}^{3/2} \varepsilon_0 H_0 \quad (120)$$

Inserting this into Equation (102) gives

$$\Delta H_0 \approx (1/6)\sqrt{\Omega_{M0}} a_{re}^{3/2} \varepsilon_0 H_0 \quad (121)$$

Using the values from ref. [16], $\Omega_{M0} = 0.3$, $a_{re} = 10^{-3}$, $\varepsilon_0 = 10^{-5}$, gives $\Delta H_0 = 3.2 \cdot 10^{-11} H_0$. This is a factor $3.2 \cdot 10^{-10}$ smaller than that needed to solve the Hubble tension. Hence, when the formulae of ref. [16] is applied to the Hubble tension, we find that the anisotropy of the expansion velocity, as deduced from the Planck data, is less than a hundred million times too small to solve the Hubble tension for this universe model.

I have a question from a referee: In Equation (112), only a small deviation from isotropy along one axis is considered. Will adding more general perturbations along all axes help to at least alleviate the Hubble tension? In order to answer this question I shall now present the above result in a different way.

According to Equation (112)

$$a_1 = a(1+\varepsilon) \,, \, a_2 = a_3 = a, \quad (122)$$

where $|\varepsilon| \ll 1$. Hence

$$H_1 \approx H + \dot{\varepsilon} \,, \, H_2 = H_3 = H . \quad (123)$$

It then follows from Equation (6) that for this universe model

$$\sigma^2 \approx \frac{1}{3}\dot{\varepsilon}^2. \tag{124}$$

Equation (24) then gives the anisotropy parameter for this universe model,

$$\Omega_{\sigma\varepsilon} = \frac{1}{9}\frac{\dot{\varepsilon}^2}{H^2 V^2}. \tag{125}$$

With the normalization $V(t_0) = 1$, the present value of the anisotropy parameter is

$$\Omega_{\sigma\varepsilon 0} = \frac{1}{9}\frac{\dot{\varepsilon}_0^2}{H_0^2}. \tag{126}$$

It follows from Equations (115) and (125) that in a universe model with only a small deviation (perturbation) from isotropy in one direction, the present magnitude of the anisotropy parameter is

$$\Omega_{\sigma\varepsilon 0} = \frac{1}{2}\left(\frac{\Delta H_0}{H_0}\right)^2, \tag{127}$$

where $\Delta H_0$ is the difference between the Hubble constant in this universe model with anisotropic expansion in one direction and in a universe with isotropic expansion.

In order for the expansion anisotropy to have a possibility of solving the Hubble tension, $\Delta H_0$ must be at least as large as the difference between the early type and late type determinations of the Hubble constant, $\Delta H_0 > 5$ km/s per Mpc. Hence, we obtain $\Omega_{\sigma\varepsilon 0} > 5 \cdot 10^{-3}$.

This result is in conflict with that of ref. [5]. This may be a result of the special character of the universe model used to deduce Equation (120)—A model with anisotropy in only one direction, i.e., a model with minimal deviation from the isotropic models. Hence, in order to shed new light upon the question of whether the expansion anisotropy can solve the Hubble tension, I shall, in the next Section, consider this question from a new point of view and use the exact solution of Einstein's field equations given in Section 2 with arbitrarily large anisotropy in all three directions to calculate how much the expansion anisotropy changes the value of the Hubble constant.

**5. The Hubble Tension Analysed by Means of the Anisotropic ΛCDM Universe Model**

The value of the Hubble constant, as determined from measurements of the temperature fluctuations in the microwave background radiation by the Planck team, is $H_0$ = (67.4 ± 0.5) km s$^{-1}$ Mpc$^{-1}$ = $1/1.45 \cdot 10^{10}$ years. With the above values of the density parameters and the anisotropy parameter, we obtain $t_\Lambda = 1.25 \cdot 10^{10}$ years and $K_\sigma = 2.8 \cdot 10^{-9}$.

We shall now use the exact solution (71) of Einstein's field equations and the corresponding expression (73) for the Hubble parameter to investigate whether the expansion isotropy is large enough to solve the Hubble tension. We define the difference between the Hubble constant in an isotropic- and anisotropic universe as $\Delta H_0 = H_0 - H_{0iso}$. In order to solve the Hubble tension, the present value of $\Delta H_0$ must be at least as large as the difference between the late time and early time measurements of the Hubble constant, $\Delta H_0 > 5$ (km/s)Mpc$^{-1}$.

The method takes as a point of departure that the value of the Hubble parameter determined from the temperature fluctuations 380,000 years after the Big Bang, at the time of recombination, $\hat{t}_{ISORC} = 3.3 \cdot 10^{-5}$, is one and the same whether the universe is assumed to be isotropic or anisotropic. It is a model-independent quantity determined directly from observations. Hence, putting $H(t_{ISORC}) = H_{ISO}(t_{ISORC})$ in Equations (39) and (73), we obtain

$$H_0 = \frac{1 + 2K_\sigma \coth \hat{t}_{ISORC}}{1 + 2K_\sigma \coth(2\hat{t}_{ISORC})} H_{\Lambda CDM0} \quad (128)$$

This is the relationship between the Hubble constant in the anisotropic- and the isotropic $\Lambda CDM$-universe as determined from the Planck measurements determining the value of the Hubble parameter at the point of time $t_{RC}$. Hence, the difference between the Hubble constant in an isotropic- and anisotropic $\Lambda CDM-$ universe can be expressed as

$$\Delta H_0 = H_0 - H_{\Lambda CDM0} = \frac{2K_\sigma}{\sinh(2\hat{t}_{ISORC}) + 2K_\sigma \cosh(2\hat{t}_{ISORC})} H_{\Lambda CDM0}. \quad (129)$$

Compared with Equation (76), this may be written as

$$\Delta H_0 = H_{\Lambda CDM0} \sqrt{2\Omega_\sigma(t_{ISORC})}. \quad (130)$$

It was shown above that $\Omega_\sigma(t_{ISORC}) = 8.3 \cdot 10^{-9}$. Hence, we obtain $\Delta H_0 = 6.7 \cdot 10^{-5} H_{\Lambda CDM0}$. This is too small to be able to solve the Hubble tension.

In appendix B it is shown that if we make a series expansion of the expression (31) for the Hubble parameter to 1. order in $\Omega_{\sigma 0}$ and then calculates $\Delta H_0$, the result is that $\Delta H_0$ seems to be proportional to $\Omega_{\sigma 0}$. However, eq.(129) and the expression for $K_\sigma$ in eq.(70), shows that $\Delta H_0$ is proportional to the square root of $\Omega_{\sigma 0}$.

It is seen from Equation (128) that the value of $\Delta H_0$ is determined by the amount of anisotropy and the recombination time, $\hat{t}_{RC}$. The earlier this is determined to happen, the closer will the cosmic anisotropy be to solving the Hubble tension. We shall now consider how much $\hat{t}_{RC}$ depends upon the anisotropy and radiation contents of the universe.

Since $\hat{t}_{ISORC} = 3.3 \cdot 10^{-5}$ is very small and $\hat{t}_{RC}$ in the anisotropic universe is assumed to be of the same order of magnitude, we can, with good accuracy, use the approximations $\sinh \hat{t}_{RC} \approx \hat{t}_{RC}$, $\cosh \hat{t}_{RC} \approx 1$ in Equation (71) and $\sinh \hat{t}_{ISORC} \approx \hat{t}_{ISORC}$ Equation (33). This gives

$$V(t_{RC}) \approx \frac{\Omega_{M0}}{\Omega_{\Lambda 0}} \left( \hat{t}_{RC}^2 + 2K_\sigma \hat{t}_{RC} \right) \approx \frac{\Omega_{M0}}{\Omega_{\Lambda 0}} \hat{t}_{ISORC}^2. \quad (131)$$

Solving this Equation with respect to $\hat{t}_{RC}$ gives

$$\hat{t}_{RC} \approx \sqrt{\hat{t}_{ISORC}^2 + K_\sigma^2} - K_\sigma \approx \hat{t}_{ISORC} - K_\sigma. \quad (132)$$

Hence, $\hat{t}_{ISORC} - \hat{t}_{RC} \approx K_\sigma$, so $\hat{t}_{RC}$ is very close to $\hat{t}_{ISORC}$. This means that the change of the recombination time due to anisotropy does not give a significant change of $\Delta H_0$ as given in Equation (129).

Next, we consider whether the presence of radiation is of greater importance in this connection. From Equation (31), it follows that the dimensionless point of time of the cosmic recombination for an isotropic, flat universe with dust, radiation, and LIVE is

$$\hat{t}_{ISORRC} = \frac{3}{2}\sqrt{\Omega_{\Lambda 0}} \int_0^{a_{RC}} \frac{a\,da}{\sqrt{\Omega_{\Lambda 0}a^4 + \Omega_{M0}a + \Omega_{R0}}}, \quad (133)$$

The last scattering redshift is given by Akarsu et al. [6] as $z_{RC} = 1090$. This corresponds to $a_{RC} = 9.17 \cdot 10^{-4}$. Using this together with the standard values of the ΛCDM -universe, $\Omega_{\Lambda 0} = 0.7$, $\Omega_{M0} = 0.3$, and $\Omega_{R0} = 0$, a numerical calculation of the integral (132) without radiation gives $\hat{t}_{ISORC} = 4.2 \cdot 10^{-5}$ corresponding to $t_{ISORC} = 4.7 \cdot 10^5$ years in agreement of the result from Equation (27). Inserting this into Equation (129) with $K_\sigma = 2.8 \cdot 10^{-9}$ gives $\Delta H_0 = 4.7 \cdot 10^{-3}$ km/s per Mpc.

Including the CMB-radiation, $\Omega_{r0} = 5 \cdot 10^{-4}$, Equation (128) gives $\hat{t}_{ISORRC} = 3.2 \cdot 10^{-5}$ corresponding to $t_{ISORRC} = 3.7 \cdot 10^5$ years, which is the standard value of the recombination time. This gives $\Delta H_0 = 6.1 \cdot 10^{-3}$ km/s per Mpc. It is still too small to have any potential for solving the Hubble tension.

I will finally answer two questions that I have received from one of the referees: 1. What level of anisotropy would be necessary to solve the Hubble tension? 2. What other observations would then be contradicted?

Concerning the first question, it follows from Equation (123) that in order to have a possibility of solving the Hubble tension, the value of the anisotropy parameter at the recombination time must obey

$$\Omega_\sigma(t_{RC}) > \frac{1}{2}\left(\frac{\Delta H_0}{H_0}\right)^2, \quad (134)$$

where $\Delta H_0$ is the average difference between the late time and early time determinations of the Hubble constant, $\Delta H_0 = 5$ (km/s)Mpc$^{-1}$. Using the value $H_0 = 70$ (km/s)Mpc$^{-1}$ for the Hubble constant, we obtain $\Omega_\sigma(t_{RC}) = 2.5 \cdot 10^{-3}$. Using Equation (76) this requires that the present value of the anisotropy parameter is

$$\Omega_{\sigma 0} > \left(\frac{\cosh(2\hat{t}_{RC}) + (1/2K_\sigma)\sinh(2\hat{t}_{RC})}{\cosh(2\hat{t}_0) + (1/2K_\sigma)\sinh(2\hat{t}_0)}\right)^2 \Omega(t_{RC}). \quad (135)$$

Here $\hat{t}_{RC} = 4.2 \cdot 10^{-5}$, $\hat{t}_0 = 1.1$, and $K_\sigma = 2.8 \cdot 10^{-9}$. Hence, the last terms dominate both in the numerator and the denominator, and $\sinh\hat{t}_{RC} \approx \hat{t}_{RC}$. Hence, we can, with good approximation, write

$$\Omega_{\sigma 0} > 2\left[\frac{\hat{t}_{RC}}{\sinh(2\hat{t}_0)} \frac{\Delta H_0}{H_0}\right]^2. \quad (136)$$

Hence, so that the expansion anisotropy should be able to solve the Hubble tension, the present value of the anisotropy parameter must obey $\Omega_{\sigma 0} > 9 \cdot 10^{-13}$.

This value of the anisotropy parameter is in conflict with several measurements of the properties of the early universe. I will here refer to the analysis of Akarsu and co-workers [6,7]. From observations of baryon acoustic oscillations (BAO) and the Planck measurements of CMB temperature fluctuations, they found the following restriction on the present value of the anisotropy parameter: $\Omega_{\sigma 0} < 10^{-15}$. Hence the values $\Omega_{\sigma 0} > 9 \cdot 10^{-13}$ are in conflict with these measurements.

Furthermore, they considered the restriction upon the expansion anisotropy coming from the Big Bang nucleosynthesis (BBN). It follows from Equations (24) and (27) that

$$\Omega_\sigma = \left(\frac{H_0}{H}\right)^2 \frac{1}{a^6}\Omega_{\sigma 0} \quad , \quad \Omega_R = \left(\frac{H_0}{H}\right)^2 \frac{1}{a^4}\Omega_{R0} \tag{137}$$

From this and Equation (3) for the redshift, we have

$$\frac{\Omega_\sigma}{\Omega_R} = (1+z)^2 \frac{\Omega_{\sigma 0}}{\Omega_{R0}}. \tag{138}$$

The calculations of the amounts of the lightest element from the BBN in a universe with isotropic cosmic expansion are in agreement with observations. A large anisotropy would change the calculated quantities and would hence give calculated amounts of helium and lithium conflicting with observations. In order to avoid such a conflict, Akarsu et al. required that the anisotropy parameter at the point of time of the BBN, around three minutes after the Big Bang, should not be larger than the radiation density at the same time, $\Omega_\sigma(t_{BBN}) < \Omega_R(t_{BBN})$. One can, with good accuracy, use the approximation $1 + z_{BBN} \to z_{BNN}$. Hence, they concluded that in order not to destroy the agreement of the predictions of the BBN calculations with observations, the present anisotropy parameter is restricted to values obeying

$$\Omega_{\sigma 0} < \frac{\Omega_{R0}}{z_{BBN}^2}. \tag{139}$$

Inserting $\Omega_{R0} = 5 \cdot 10^{-4}$ and $z_{BBN} = 3 \cdot 10^8$ gives $\Omega_{\sigma 0} < 5.6 \cdot 10^{-21}$. This means that the values $\Omega_{\sigma 0} > 9 \cdot 10^{-13}$ needed to solve the Hubble tension would lead to a severe conflict between BBN calculations in such an anisotropic universe and the observed cosmic amounts of the lightest element.

In Section 4, a universe model with only a small deviation from isotropy along one axis was considered. We can now answer the question: "Will adding more general perturbations along all axes help to at least alleviate the Hubble tension?" It was found that for a universe model with only a small deviation from isotropy along one axis, the present value of the expansion anisotropy parameter must be rather large, $\Omega_{\alpha 0} > 5 \cdot 10^{-3}$, in order for the anisotropy shall be able to solve the Hubble tension. In the context of a universe model with anisotropic expansion in all three directions it is sufficient that $\Omega_{\sigma 0} > 9 \cdot 10^{-13}$. Therefore, "adding more general perturbations along all axes" does indeed alleviate the Hubble tension', in fact, by 10 orders of magnitude. But even that is not sufficient to solve the Hubble tension.

## 6. Conclusions

The main topic of the present paper has been the presentation of the anisotropic Bianchi type I generalization of the isotropic ΛCDM-universe model. Considering the Equations (59)–(80), the evolution of this universe model can be separated into three periods. (1) Initially, the universe was in a Kasner-like era, where anisotropy dominated over matter and LIVE. The universe came from a singular state with vanishing co-moving volume inside the Hubble horizon and infinitely great Hubble parameter. The anisotropy parameter originally had its maximal value $\Omega_\sigma = 1$. In this period, the deceleration parameter vanished. (2) There was a transition from an early anisotropy-dominated period to a matter-dominated era at the point of time $t_1$ given in Equation (73). The recombination time when the universe became transparent to the CMB-radiation took place early in this era. (3) Transition from a matter-dominated era with cosmic deceleration due to the attractive gravity of the matter to an era with repul-sive gravity of the LIVE took place at a point of time given in Equation (34). The predicted redshift at the transition is $z_{TR} = 0.67$.

It has recently been argued that anisotropy of the universe can solve the Hubble tension. In this connection, the Bianchi type I models are of unique importance because they are the only Bianchi models that contain the ΛCDM universe model as a special case. Hence, they are the proper generalization of our standard model of the universe.

Einstein's field equations have therefore here been solved for a general Bianchi type I universe with cold matter and LIVE with a constant density, which can be represented by a cosmological constant, and the most general model of this type has been presented in a way suitable for investigations of the effects of expansion anisotropy upon observable properties of the models. In particular two models have been applied to an investigation of the Hubble tension, one with deviation from isotropic expansion in only one direction, and a general Bianchi type I model.

In this way, I have shown that the expansion anisotropy of the universe model with deviation from isotropy in only one direction is much too small to solve the Hubble tension. In the case of the general Bianchi type I, a calculation based upon the exact solution of Einstein's field equations for a general Bianchi type I universe with dust and LIVE leads to a larger effect of the expansion anisotropy upon the value of the Hubble constant than in the previous cases. However, even if radiation is included in the universe model when the point of time of the recombination is calculated, and the constraints come from the requirement that the cosmic expansion anisotropy shall not have a significant effect upon the cosmic nucleosynthesis during the first ten minutes of the history of the universe is neglected, the permitted anisotropy is still not large enough to solve the Hubble tension.

Akarsu and co-workers have shown that if one takes into account the constraint that the expansion anisotropy shall have no significant effect on the standard Big Bang nucleosynthesis, the largest permitted value of the anisotropy parameter is $\Omega_{\sigma 0} = 10^{-23}$ [6], which is a hundred thousand times smaller than if these constraints are neglected. Hence, if this last requirement is not neglected, the permitted anisotropy is much too small to solve the Hubble tension.

**Funding:** This research received no external funding.

**Data Availability Statement:** Data are contained within the article.

**Acknowledgments:** I thank the referees for constructive suggestions that contributed to improvements of the article.

**Conflicts of Interest:** The author declares no conflict of interest.

## Appendix A. Determination of the Transition Time from Deceleration to Acceleration in the Anisotropic Universe

The transition time $t_2$ from deceleration to acceleration in the anisotropic universe is given with sufficient accuracy by inserting $q(t_2) = 0$ in Equation (74). Hence,

$$q(\hat{t}_2) = q_{\Lambda CDM}(\hat{t}_2) - \frac{3K_\sigma}{\sinh \hat{t}_2 \cosh \hat{t}_2} = 0, \tag{A1}$$

giving

$$\sinh \hat{t}_2 \cosh \hat{t}_2 = \frac{3K_\sigma}{q_{\Lambda CDM}(\hat{t}_2)}. \tag{A2}$$

Since the effect of the expansion anisotropy is small, we can, with good accuracy, use a series expansion of the deceleration parameter about the point of time $t_{\Lambda CDM2}$ to first order in the change, $\Delta t_2$, of the transition time $t_2 = t_{\Lambda CDM2} + \Delta t_2$ from deceleration to acceleration due to the expansion anisotropy, in order to simplify the calculation of $\Delta t_2$. This gives

$$q_{\Lambda CDM}(\hat{t}_2) = q_{\Lambda CDM}(\hat{t}_{\Lambda CDM2} + \Delta \hat{t}_2)$$
$$\approx q_{\Lambda CDM}(\hat{t}_{\Lambda CDM2}) + q'(\hat{t}_{\Lambda CDM2})\Delta \hat{t}_2 = 0 - 3\frac{\sinh \hat{t}_{\Lambda CDM2}}{\cosh^3 \hat{t}_{\Lambda CDM2}}\Delta \hat{t}_2 \quad (A3)$$

From Equation (41), we have

$$\cosh \hat{t}_{\Lambda CDM2} = \sqrt{\frac{3}{2}} \quad , \quad \sinh \hat{t}_{\Lambda CDM2} = \sqrt{\frac{1}{2}} \quad (A4)$$

giving

$$q_{\Lambda CDM}(\hat{t}_2) \approx \frac{2}{\sqrt{3}}\Delta \hat{t}_2 \approx 1.15 \Delta \hat{t}_2. \quad (A5)$$

Furthermore,

$$\cosh \hat{t}_2 = \cosh(\hat{t}_{\Lambda CDM2} + \Delta \hat{t}_2) = \cosh \hat{t}_{\Lambda CDM2} \cosh \Delta \hat{t}_2 - \sinh \hat{t}_{\Lambda CDM2} \sinh \Delta \hat{t}_2$$
$$\approx \cosh \hat{t}_{\Lambda CDM2} - \sinh \hat{t}_{\Lambda CDM2} \Delta \hat{t}_2 \quad (A6)$$

Hence,

$$\cosh \hat{t}_2 \approx \sqrt{\frac{1}{2}}\left(\sqrt{3} - \Delta \hat{t}_2\right) \quad , \quad \sinh \hat{t}_2 \approx \sqrt{\frac{1}{2} - \sqrt{3}\Delta \hat{t}_2} \,. \quad (A7)$$

Inserting Equations (A5) and (A7) into Equation (A2) gives to first order in $\Delta \hat{t}_2$

$$\Delta \hat{t}_2 \approx \frac{4}{\sqrt{3}}\frac{K_\sigma}{1.15} \approx 2.6 K_\sigma = 1.4 \cdot 10^{-6}. \quad (A8)$$

This gives $\Delta t_2 = 1.4 \cdot 10^{-6} t_\Lambda = 1.5 \cdot 10^4$ years.

**Appendix B. Significance of a Universe Model, Which Is an Exact Solution of Einstein's Field Equations**

In order to illustrate the point of the heading of this Appendix, it is useful to consider the anisotropic extension of the $\Lambda CDM-$ universe, i.e., a flat universe with LIVE and dust, neglecting radiation, since the field equations can be solved exactly in terms of elementary functions for such a model.

In this case, Equation (31) takes the form

$$H = \frac{\sqrt{\Omega_{\Lambda 0}V^2 + \Omega_{M0}V + \Omega_{\sigma 0}}}{V}H_0. \quad (A9)$$

Hence,

$$H_{iso} = \frac{\sqrt{\Omega_{\Lambda 0}V^2 + \Omega_{M0}V}}{V}H_{0iso}. \quad (A10)$$

Putting $H(a_1) = H_{iso}(a_1) = H_{measured}$ in Equations (A9) and (A10) we obtain,

$$H_0 = \frac{\sqrt{\Omega_{\Lambda 0}V_1^2 + \Omega_{M0}V_1}}{\sqrt{\Omega_{\Lambda 0}V_1^2 + \Omega_{M0}V_1 + \Omega_{\sigma 0}}}H_{0iso}, \quad (A11)$$

which may be written

$$H_0 = \frac{H_{0iso}}{\sqrt{1 + \frac{\Omega_{\sigma 0}}{\Omega_{\Lambda 0}V_1^2 + \Omega_{M0}V_1}}} \quad (A12)$$

Now $\Omega_{\sigma 0} \simeq 4 \cdot 10^{-14}$ and $a_1 \sim 10^3$, i.e., $V_1 = a_1^3 \sim 10^{-9}$ and $\Omega_{M0} \sim 0.3$. Hence, the last term inside the square root is much less than one, and the first term in the denominator is much less than the second one. We can, therefore, with good approximation, make a series expansion to first order in $\Omega_{\sigma 0}$ in the expression (A12) for $H_0$ and neglect the first term in the denominator. This gives

$$H_0 \approx H_{0iso} - \frac{\Omega_{\sigma 0}}{2\Omega_{M0}V_1}H_{0iso} \quad (A13)$$

Thus, the difference in the value of the Hubble constant with and without expansion anisotropy is

$$\Delta H_0 = H_{0iso} - H_0 \approx \frac{\Omega_{\sigma 0}}{2\Omega_{M0}V_1}H_{0iso}, \quad (A14)$$

giving $\Delta H_0 = 1.7 \cdot 10^{-9} H_{0iso}$. This is much less than that of Equation (129). There is a conflict between this result and that of Equation (129).

The solution of this conflict is hidden in the mathematical properties of the exact solution of Einstein's field equations which Equation (129) is deduced from. Using expression (71) for *V* gives

$$\sqrt{\Omega_{\Lambda 0}V^2 + \Omega_{M0}V + \Omega_{\sigma 0}} = \frac{1}{2}\frac{\Omega_{M0}}{\sqrt{\Omega_{\Lambda 0}}}\sinh(2\hat{t}) + \sqrt{\Omega_{\sigma 0}}\cosh(2\hat{t}). \quad (A15)$$

Hence, Equations (4) and (31) lead to the expression (72) for the Hubble parameter. This can be written in the form (73) and leads to the exact expression (129) for $\Delta H_0$, showing that $\Delta H_0$ depends upon $\sqrt{\Omega_{\sigma 0}}$ and not $\Omega_{\sigma 0}$ as it looks like in Equation (A14). This demonstrates that a universe model, which is an exact solution of Einstein's field equations, is useful for analyzing observable properties of the model, for example, whether the anisotropy of the cosmic expansion can solve the Hubble tension.